\documentclass[12pt,A4paper]{article}
\usepackage{graphics}
\usepackage{amsmath}
\usepackage{epsfig,amssymb}
\usepackage{graphicx}

\usepackage{color}

\begin{document}

\title{\bf Intermittent origin of  the large violations of the fluctuation
dissipation relations in an aging polymer glass}
\author{L. Buisson, S. Ciliberto, A. Garcimart\'{\i}n \thanks{%
Permanent address: Departamento de F\'{\i}sica, Facultad de
Ciencias,
Universidad de Navarra, E-31080 Pamplona, Spain.}  \\
Ecole Normale Sup\'erieure de Lyon, Laboratoire de Physique ,\\
C.N.R.S. UMR5672,  \\ 46, All\'ee d'Italie, 69364 Lyon Cedex 07,
France\\}

\maketitle

\begin{abstract}
The fluctuation-dissipation relation (FDR) is measured on the
dielectric properties of a polymer glass (polycarbonate)in the
range $20mHz - 100Hz$. It is found that after a quench below the
glass transition temperature the fluctuation dissipation theorem
is strongly violated. The amplitude and the persistence time of
this violation are decreasing functions of frequency. At
frequencies larger than 1Hz  it persists for about $3h$. The
origin of this violation is a highly intermittent dynamics
characterized by large fluctuations. The relevance of these
results for recent models of aging dynamics are discussed.
\end{abstract}


\bigskip


 When glassy materials are quenched from above their glass
transition temperature $T_g$ to a temperature lower than $T_g$,
any response function of these materials depends on the time $t_w$
elapsed from the quench: they are aging \cite{Struick}.
A widely studied question, is how the temperature of these systems
can be defined. Recent theories \cite{Kurchan} based on the
description of spin glasses by a mean field approach propose to
extend the concept of temperature using a Fluctuation Dissipation
Relation (FDR) which generalizes for a weakly out of equilibrium
system the Fluctuation Dissipation Theorem (FDT) . (For a review
see ref. \cite{Bouchaud,Peliti}).
At equilibrium, FDT relates the fluctuation spectral density
 of a variable V to the response $\chi_{Vq} (f)$
 of V to a perturbation of its
conjugated variable q at frequency f:
\begin{equation}
S(f) = { 2 K_B \ T \over \pi f } {\it Im}\left[\chi_{Vq}(f)
\right] \label{FDR}
\end{equation}
where $S(f)=<|V(f)|^2>$ is the fluctuation spectral density of
$V$, $K_B$ is the Boltzmann constant, T the temperature of the
system and ${\it Im}\left[ \chi_{Vq}(f) \right]$ is the imaginary
part of $\chi_{Vq}(f)$. When the system is not in equilibrium FDT,
that is eq.\ref{FDR}, may fail.
Because of the slow dependence on $t_w$ of the response functions,
it has been proposed to use an FDR which generalizes  eq.\ref{FDR}
and which can be used to define
  an effective temperature $T_{eff}(f,t_{w})$ of the system \cite{Peliti}:
\begin{equation}
T_{eff} (f,t_w) = { S(f,t_w) \ \pi f \over  {\it Im}\left[
\chi_{Vq}(f,t_w) \right] \ 2 K_B }
 \label{Teff}
\end{equation}
It is clear that if eq.\ref{FDR} is satisfied $T_{eff}=T$,
otherwise $T_{eff}$ turns out to be a decreasing function of $t_w$
and $f$. The physical meaning of eq.\ref{Teff} is that there is a
time scale (for example $t_w$) which allows to separate the fast
processes from the slow ones. In other words the low frequency
modes, such that $f t_w <1$,  relax towards equilibrium much
slower than the high frequency ones which rapidly relax to the
temperature of the thermal bath.
This striking behavior has been observed in several numerical
models of aging \cite{Kob}-\cite{Marinari}, which show
that eq.\ref{Teff} is a good definition of temperature in the
thermodynamic sense \cite{Peliti}.
The experimental analysis of the dependence of $T_{eff}(f,t_w)$ on
$f$ and $t_w$ is very useful to distinguish among different models
of aging because the FDT violations are model dependent
(\cite{Peliti}-\cite{Marinari}).

 Recently, a few experiments have
analyzed this problems  in real materials \cite{Israeloff},
\cite{Bellon}, \cite{Herisson}. The violation of FDT measured in
an experiment  on a spin glass \cite{Herisson} seems to be in
agreement with theoretical
predictions. 
The  experiment  on the  dielectric measurements on glycerol
\cite{Israeloff} is a single frequency experiment. Thus cannot
give insight on the time evolution of the spectrum. The experiment
on  colloidal glasses\cite{Bellon} presents only a qualitative
agreement with theory. Specifically it  shows that the persistence
time of the violation is very long and the amplitude of the
violation is huge.


In order to give more insight into this problem  we have done wide
band ($20mHz,100Hz$) measurements of the dielectric susceptibility
and of  the polarization noise in a polymer glass.
 In
this letter we present results which show  a strong violation of
the FDT in a polymer glass quenched from the molten state to below
its glass-transition temperature. $T_{eff}$ defined by
eq.\ref{Teff} slowly relaxes towards the bath temperature. The
violation is observed even at $f t_w >> 1$ and it  may last for
more than $3h$ for $f>1Hz$.
 The polymer
used is Makrofol DE 1-1 C, a bisphenol A polycarbonate, with $T_g
\simeq 419K$, produced by Bayer in form of foils. This material
has been chosen because it has a wide temperature range of strong
aging \cite{Struick} and is totally amorphous.
Many studies of the dielectric susceptibility of this material
exist, but no one had an interest on the problem of noise
measurements.
\begin{figure}
  \centerline{\epsfysize=0.5\linewidth \epsffile{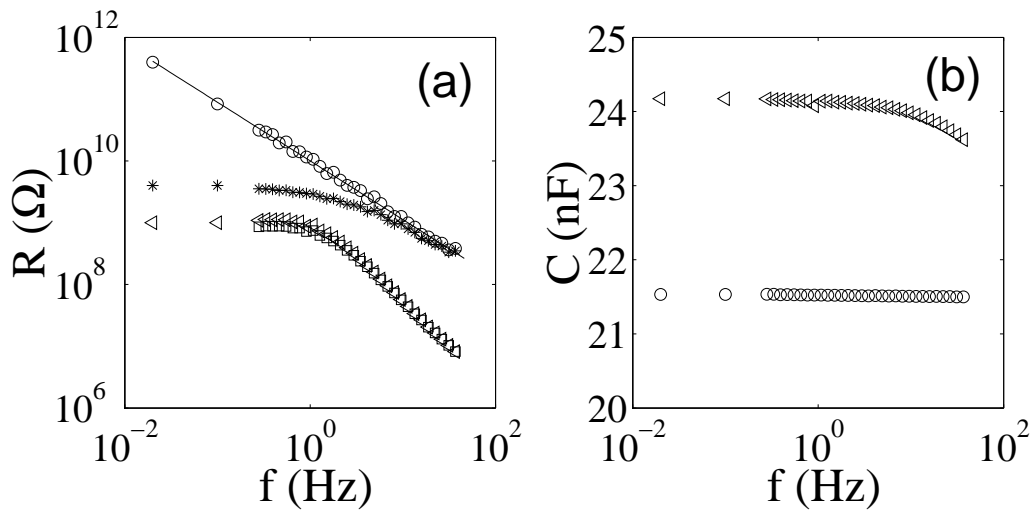}}
  \caption{(a) Polycarbonate resistance $R$  as  a function of frequency measured at $T_i=433K$ ($\vartriangleleft$) and
  at  $T_f=333K$ ($\circ$). The effect of the $4G\Omega$ input resistance is also shown at $T=433K$ ($\square$) and
   at $T=333K$ ($\ast$).
  b) Polycarbonate capacitance versus frequency measured at $T_i=433K$ ($\vartriangleleft$) and
  at  $T_f=333k$ ($\circ$).}
\label{reponse}
  \end{figure}
In our experiment polycarbonate is  used as the dielectric of a
capacitor. The  capacitor is composed by $14$ cylindrical
capacitors in parallel in order to reduce the resistance of the
sample and to increase its capacity. Each capacitor  is made of
two aluminum electrodes, $12\mu m$ thick, and  by  a disk of
polycarbonate of diameter $12cm$ and thickness $125\mu m$.
The $14$ capacitors are sandwiched together and put inside two
thick aluminum plates which contain an air circulation used  to
regulate the sample temperature within a few percent.  This
mechanical design of the capacitor is very stable and gives very
reproducible (better than 1$\%$) results even after many
temperature quenches. The capacitor is inside two Faraday screens
to insulate it from external noise. Fast quench of about $50K/min$
are obtained by injecting Nitrogen vapor in the air circulation of
the aluminum plates. The electrical impedance of the capacitor is
$Z(f,t_w) = R / (1+i \ 2 \pi \ f \ R \ C)$, where $C$ is the
capacitance and $R$ is a parallel resistance which accounts for
the complex dielectric susceptibility. It is measured using a
Novocontrol Dielectric Analyzer. The noise spectrum of this
impedance $S_Z(f,t_w)$ is:
\begin{equation}
S_Z(f,t_w)= {4 \ K_B \ T_{eff}(f, t_w) \ R \over 1+ (2 \pi \ f \ R
\ C)^2 } \label{SZ}
\end{equation}
where $T_{eff}$ is the effective temperature of the sample. The
polarization noise of $Z$ is sent to an amplifier whose input
resistance is  $R_i= 4G\Omega$. The output signal of the
amplifier is 
directly acquired by a NI4462 card. The measured spectrum at the
amplifier input is:
\begin{eqnarray}
S_V(f,t_w) &= & {4 \  K_B \ R \ R_i \ \ (\ T_{eff}(f, t_w) \ R_i +
\ T_R \ R ) \over (R+R_i)^2+(2 \pi \ f  \ R \ R_i \ C)^2} + \notag \\
 &+&  { S_\xi(f) \ R \ R_i \over (R+R_i)^2+(2 \pi \ f  \ R \ R_i \
 C)^2}+ S_{\eta}(f)
 \label{Vnoise}
\end{eqnarray}
where $T_R$ is the temperature of $R_i$ and $S_\eta$ and $S_\xi$
are respectively the voltage and the current noise spectrum of the
amplifier.
 The desired statistical accuracy  of
 $S_V(f,t_w)$ is reached by averaging the results of many
 experiments. In  each of these experiments
 the sample is first heated  to $T_i=433K$. It is maintained at this temperature
 for 4 hours to reinitialize its thermal history. Then it is quenched from $T_i$
to $T_f=333K$ in about 2 minutes. 
The origin of aging time
$t_w$ is the instant when the capacitor temperature is at $T_g
\simeq 419 K$, which of course may depend on the cooling rate.
However adjustment of $T_g$ of a few degrees will shift the time
axis by at most $30s$, without affecting our results.

At $T_f=333K$ the main relaxation process of polycarbonate are
well outside the frequency range of our measurements, specifically
the $\alpha$ relaxation  frequency is smaller than $10^{-10}Hz$
and the $\beta$ relaxation  frequency is larger than $10^{6}Hz$.
Thus we are testing a frequency range which is far away the two
relaxation frequencies.
 In fig.\ref{reponse}(a) and (b), we plot the measured values of $R$ and $C$
 as a function of $f$ at $T_i$ and at $T_f$ for $t_w \geqslant 200s$.
  We see that lowering temperature $R$ increases and $C$
decreases. At $T_f$ aging is small and  extremely slow. Thus for
$t_w>200s$ the impedance can be considered constant without
affecting our results.  From the data plotted in fig.\ref{reponse}
(a) and (b) one finds that $R=10^{10}(1 \pm 0.05) \ f^{-0.95\pm
0.01} \ \Omega$ and $C=(21.5 \pm 0.05) nF$. In
fig.\ref{reponse}(a) we also plot the total resistance at the
amplifier input which is the parallel of the capacitor impedance
with $R_i$. We see that at $T_f$ the input impedance of the
amplifier is negligible  for $f>10Hz$, whereas it has to be taken
into account at slower frequencies.

\begin{figure}[ht!]
\centerline{{\epsfxsize=0.5\linewidth\epsffile{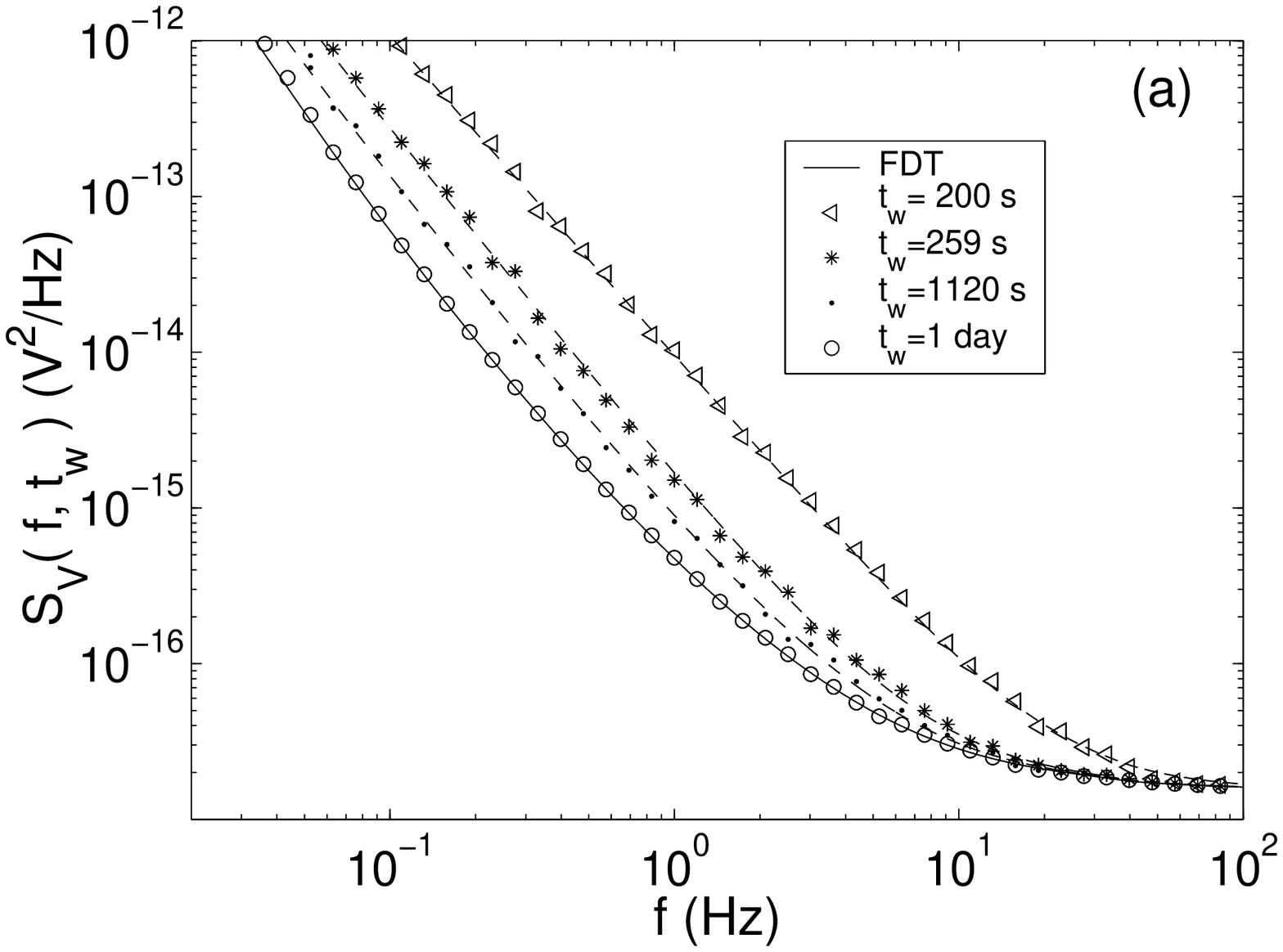}}
{\epsfxsize=0.5\linewidth\epsffile{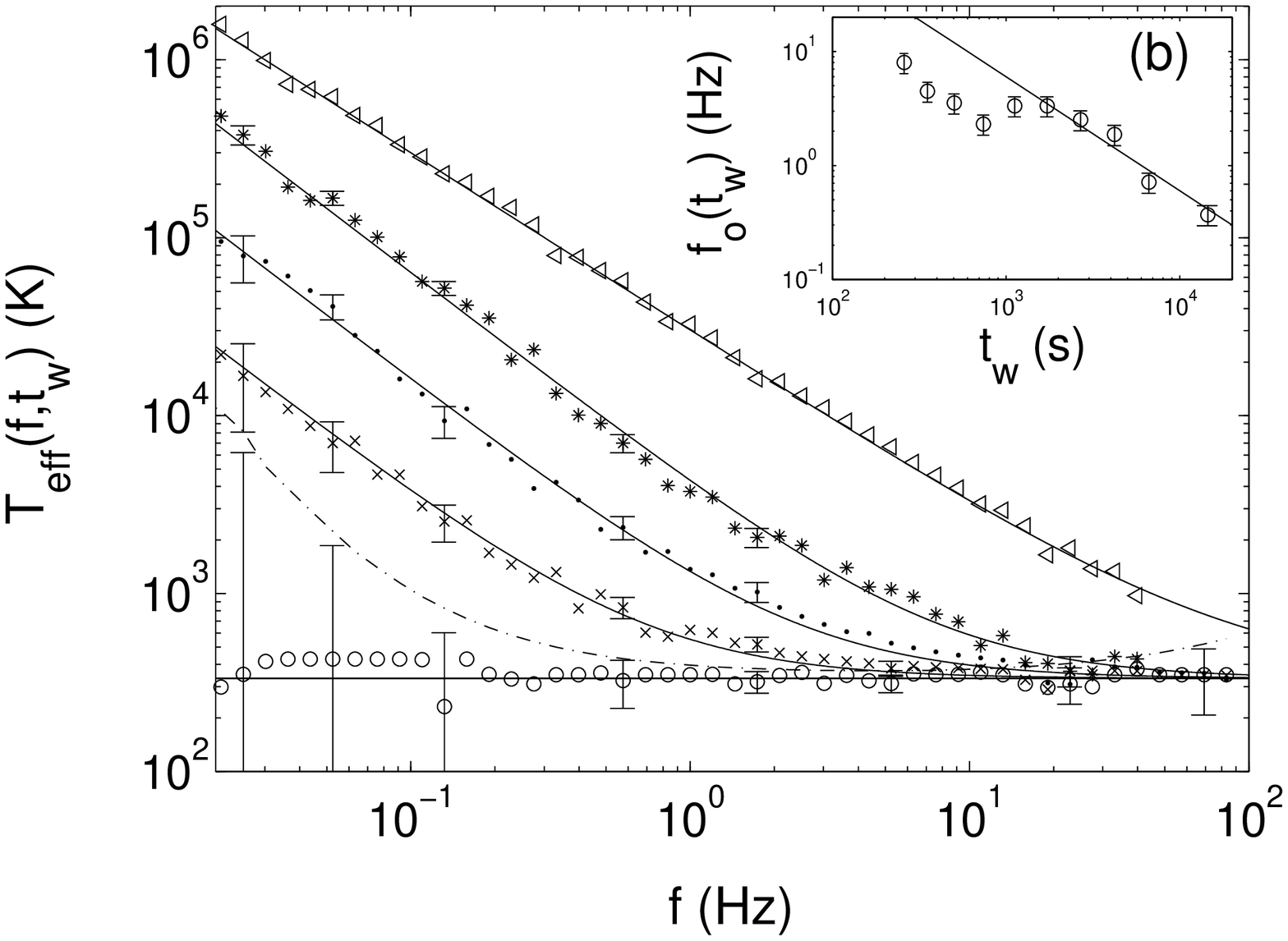}}}
  \caption{(a)  Noise power spectral density $S_V(f,t_w)$ measured at $T_f=333K$ and different
  $t_w$. The spectra are the average
over seven quenches.
  The continuous line is the FDT prediction. Dashed lines
   are the fit obtained using eq.\ref{Vnoise} and eq.\ref{fitTeff} (see text for details).
(b)  Effective temperature vs frequency at $T_f=333K$ for
different aging times: $ (\vartriangleleft)\ tw= 200 \ s$, $
(\ast)\ tw= 260 s$,  $ \bullet \ tw= 2580 s$, $ (\times)
t_w=6542s$, $ (\circ) t_w= 1\ day $. The continuous lines are the
fits obtained using eq.\ref{fitTeff}. The horizontal straight line
is the FDT prediction.  The dot dashed line corresponds to the
limit where the FDT violation can be detected. In the inset the
frequency $f_o(t_w)$, defined in eq.\ref{fitTeff},is plotted as a
function of $t_w$. The continuous line is not a fit, but it
corresponds to $f_o(t_w) \propto 1/t_w $. } \label{noise}
\end{figure}

 Fig.\ref{noise}(a) represents the evolution of
$S_V(f,t_w)$ after a quench. Each spectrum is obtained as an
average in a time window starting at $t_w$. The time window
increases with $t_w$ so to reduce errors for large $t_w$. Then the
results of 7 quenches have been averaged. At longest time ($t_w=1
day$) the equilibrium FDT prediction (continuous line) is quite
well satisfied. FDT is strongly violated for all frequencies at
short times. Then high frequencies relax on the FDT, but there is
a persistence of the violation for lower frequencies. The amount
of the violation can be estimated by the best fit  of
$T_{eff}(f,t_w)$ in eq.\ref{Vnoise} where all other parameters are
known. We start at very large $t_w$ when the system is relaxed and
$T_{eff}=T$ for all frequencies.  Inserting the values in
eq.\ref{Vnoise} and using the $S_V$ measured at $t_w=1 day$ we
find $T_{eff}\simeq 333K$, within error bars for all frequencies
(see fig.\ref{noise}b). At short $t_w$  data show that
$T_{eff}(f,t_w)\simeq T_f$ for $f$ larger than a cut-off frequency
$f_o(t_w)$ which is a function of $t_w$. In contrast, for
$f<f_o(t_w)\ \ $, 
 $T_{eff}(f,t_w)\propto
f^{-A(t_w)}$, with $A(t_w)\simeq 1$. This frequency dependence of
$T_{eff}(f,t_w)$ is quite well approximated by
\begin{equation} T_{eff}(f,t_w)= T_f \ [ \ 1 \ + \ ( {f \over
f_o(t_w)})^{-A(t_w)} \ ]
 \label{fitTeff}
\end{equation}
where $A(t_w)$ and  $f_o(t_w)$  are the  fitting parameters. We
find that $1<A(t_w)<1.2$ for all the data set. Furthermore  for
$t_w \geq 250s$, it is enough to keep $A(t_w)=1.2$ to fit the data
within error bars. For $t_w <250s$ we fixed $A(t)=1$. Thus the
only free parameter in eq.\ref{fitTeff} is  $f_o(t_w)$.  The
continuous lines in fig.\ref{noise}(a) are the best fits of $S_V$
found inserting eq.\ref{fitTeff} in eq.\ref{Vnoise}.
In fig.\ref{noise}(b) we plot the estimated $T_{eff}(f,t_w)$  as a
function of frequency at different $t_w$. Just after the quench
$T_{eff}(f,t_w)$ is much larger than $T_f$  in all the frequency
interval. High frequencies rapidly decay towards the FDT
prediction whereas  at the smallest frequencies $T_{eff}\simeq
10^5K$. Moreover low frequencies decay more slowly than high
frequencies and the evolution of $T_{eff}(f,t_w)$ towards
equilibrium is very slow.
 From the data of Fig.\ref{noise}(b) and eq.\ref{fitTeff}, it is
easy  to see that $T_{eff}(f,t_w)$ can be superposed onto a master
curve by plotting them as a function of $f/f_o(t_w)$. The function
$f_o(t_w)$ is a decreasing function of $t_w$, but the dependence
is not a simple one, as it can be seen in the inset of
fig.\ref{noise}(b). The continuous straight line is not a fit,  it
represents $f_o(t_w)\propto 1./t_w$ which seems a reasonable
approximation  for these data. For $t_w > 10^4 s$ we find the
$f_o<1Hz$. Thus we cannot follow the evolution of $T_{eff}$
anymore  because the contribution of the experimental noise on
$S_V$ is too big, as it is shown in Fig.\ref{noise}(b) by the
increasing of the error bars for $t_w=1 day$ and $f<0.1 Hz$.

Before discussing these experimental results we want to compare
them to the single frequency experiment performed on glycerol
\cite{Israeloff}. In this experiment, $T_{eff}$ has been measured
only at $7Hz$ and  nothing can be said on the dependence of
$T_{eff}$ on frequency.  This is a crucial point in order to
compare the observed dynamics to that of  models \cite{Peliti} and
to see whether there is a connection between $T_{eff}$ and the
concept of fictive temperature (see discussion in ref.
\cite{Peliti}). The strong frequency dependence of $T_{eff}$
suggest that  such a comparison will make sense only for
$f\rightarrow 0$. Therefore  it would be interesting to check
whether at shorter times and at lower frequencies large $T_{eff}$
could be observed in glycerol too.

To compare with theoretical predictions \cite{Kurchan,Peliti} and
recent spin glass experiment \cite{Herisson} we may plot the
integrated response $R(t,t_w)$ as a function of the correlation
$C(t,t_w)$. The latter is obtained inserting the measured
$T_{eff}(f,t_w)$ in eq.\ref{SZ} and Fourier transforming it.
$R(t,t_w)$ can be computed by Fourier transforming
$Real[Z(f,t_w)]$. FDR now takes the form \cite{Peliti}:
\begin{equation}
-C(t,t_w)+C(t_w,t_w)= K_B \ T_{eff}(t,t_w) \ R(t,t_w)
\label{correlation}
\end{equation}

The study of the correlations functions will be the object of a
longer reports\cite{Buisson}. Here we want only to stress  that
the selfsimilarity of $T_{eff}$ for $t_w>300s$ (see
eq.\ref{fitTeff}) implies that also $C(t_w, t)$  can be scaled
onto a single master curve by plotting $C(t_w, t)$ as a function
$(t-t_w)/t_o(t_w)$, where $t_o(t_w)$ is an increasing function of
$t_w$: approximately $t_o(t_w) \propto 1/f_o(t_w)$ for $t_w>300s$.
The self-similarity of correlation functions, found on our
dielectric data, is a characteristic of the universal picture of
aging \cite{Bouchaud,Peliti,Kob,Berthier2}, which  has been also
observed in spin-glass experiment \cite{Herisson} and in the
structure function of the dynamic light scattering of colloidal
gels \cite{Weitz}. Thus our results show  that this picture of
aging applies also to the dielectric noise of a polymer.
Furthermore the study of  $R(t,t_w)$ as a function
$(-C(t,t_w)+C(t_w,t_w))/K_B$ confirms the spectral analysis of
fig.\ref{noise}, that is $T_{eff}$ is huge for long $(t-t_w)$.
This result is quite different to what has been observed in recent
experiments on spin glasses where $T_{eff}\simeq 5 T_g$ has been
measured\cite{Herisson}.

\begin{figure}[!ht]
            \centerline{{\epsfxsize=0.5\linewidth
\epsffile{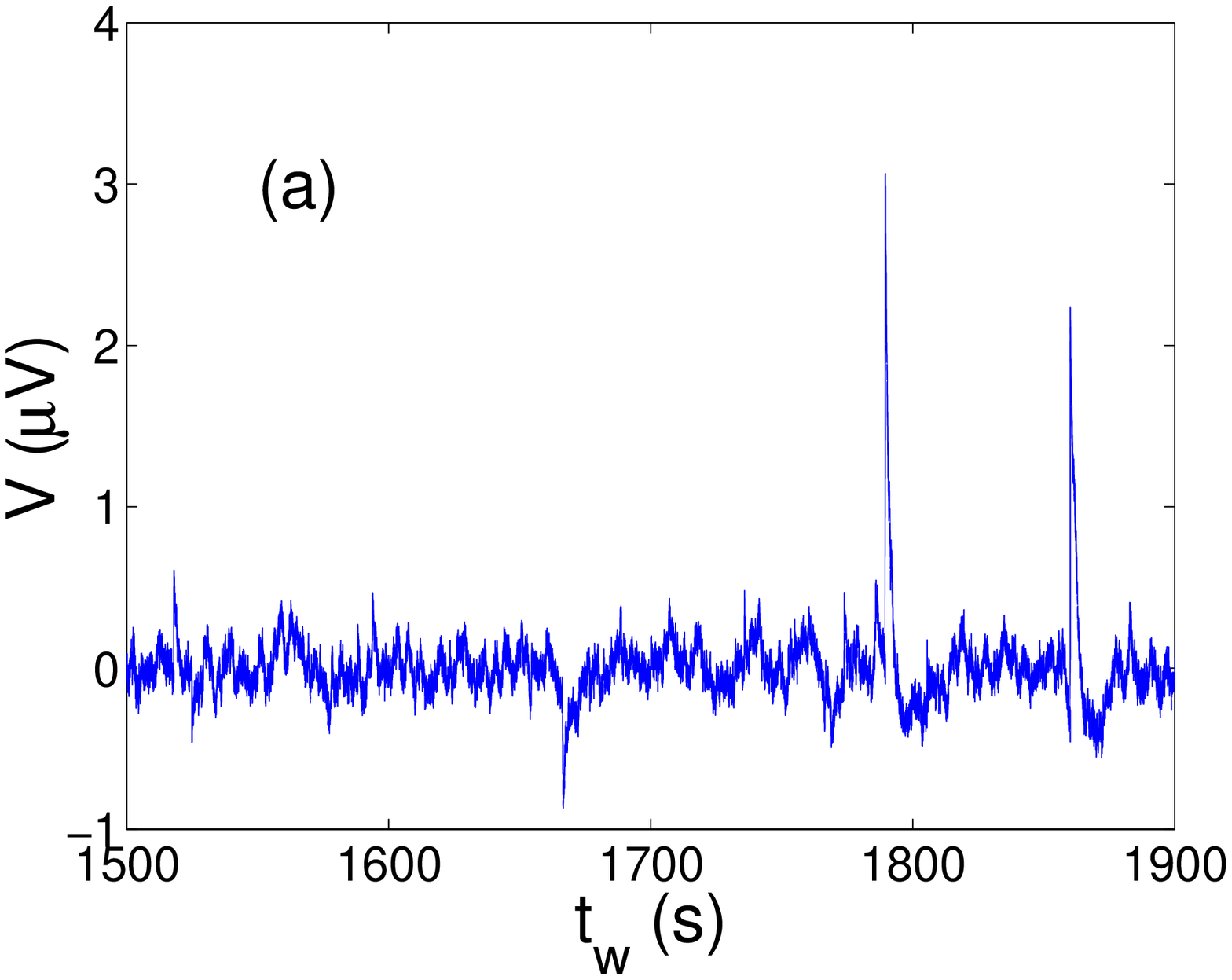}} {\epsfxsize=0.5\linewidth
\epsffile{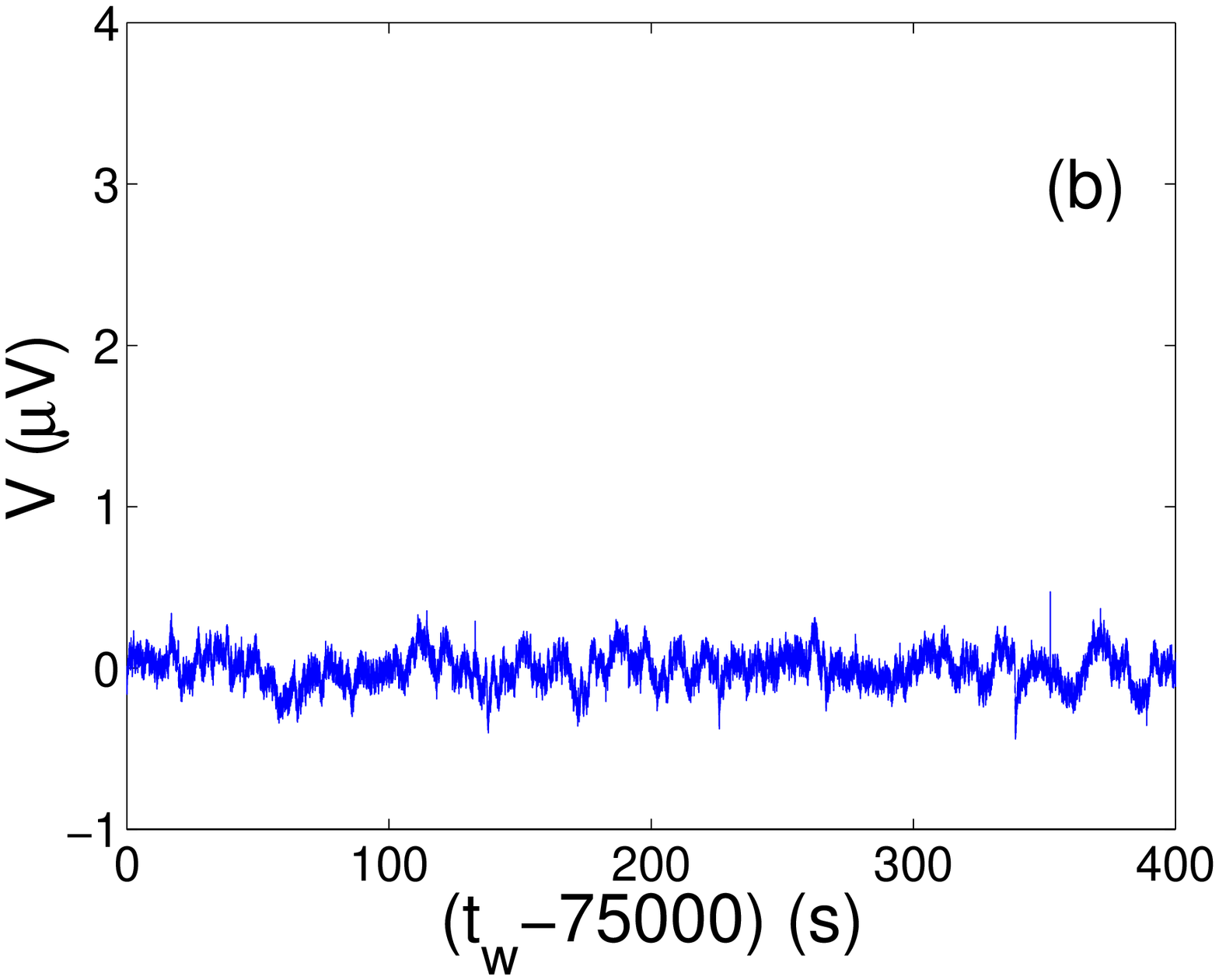}}} \caption{ {\bf Voltage
noise signal in polycarbonate} Typical noise signal of
polycarbonate measured at $1500s<t_w<1900s$ (a) and $t_w>75000s $
(b)}
 \label{signalpolyca}
\end{figure}

A large $T_{eff}$ is not specific to our system but it has been
observed in domain growth models \cite{Peliti,Barrat} and in
models controlled by activation processes
\cite{Bouchaud,trap,Sollich,Miguel}. The question is whether these
models may have some connections with our observations. In order
to understand the origin of such large deviations in our
experiment we have analyzed  the noise signal. We find that the
signal is characterized by  large intermittent events which
produce  low frequency spectra proportional to $f^{-\alpha}$ with
$\alpha \simeq 2$. Two  typical signals recorded at
$1500s<t_w<1900s$ and $t_w>75000s$  are plotted in
fig.\ref{signalpolyca}. We clearly see that in the signal recorded
at $1500s<t_w<1900s$  there are very large bursts which are on the
origin  of the frequency spectra previously. In contrast in the
signal (fig.\ref{signalpolyca}b), which was recorded at
$t_w>75000s$ when FDT is not violated, the bursts are totally
disappeared. It is interesting to notice  that this kind of
dynamics could be a common feature of several aging systems (see
also ref.\cite{Cugliandolo}). Indeed intermittency has been also
observed in dielectric local measurements of polymers
\cite{Israeloff_Nature} and in the diffusing wave spectroscopy
measurements in gels \cite{Cipelletti}.

\begin{figure}[!ht]
\centerline{{\epsfxsize=0.5\linewidth
\epsffile{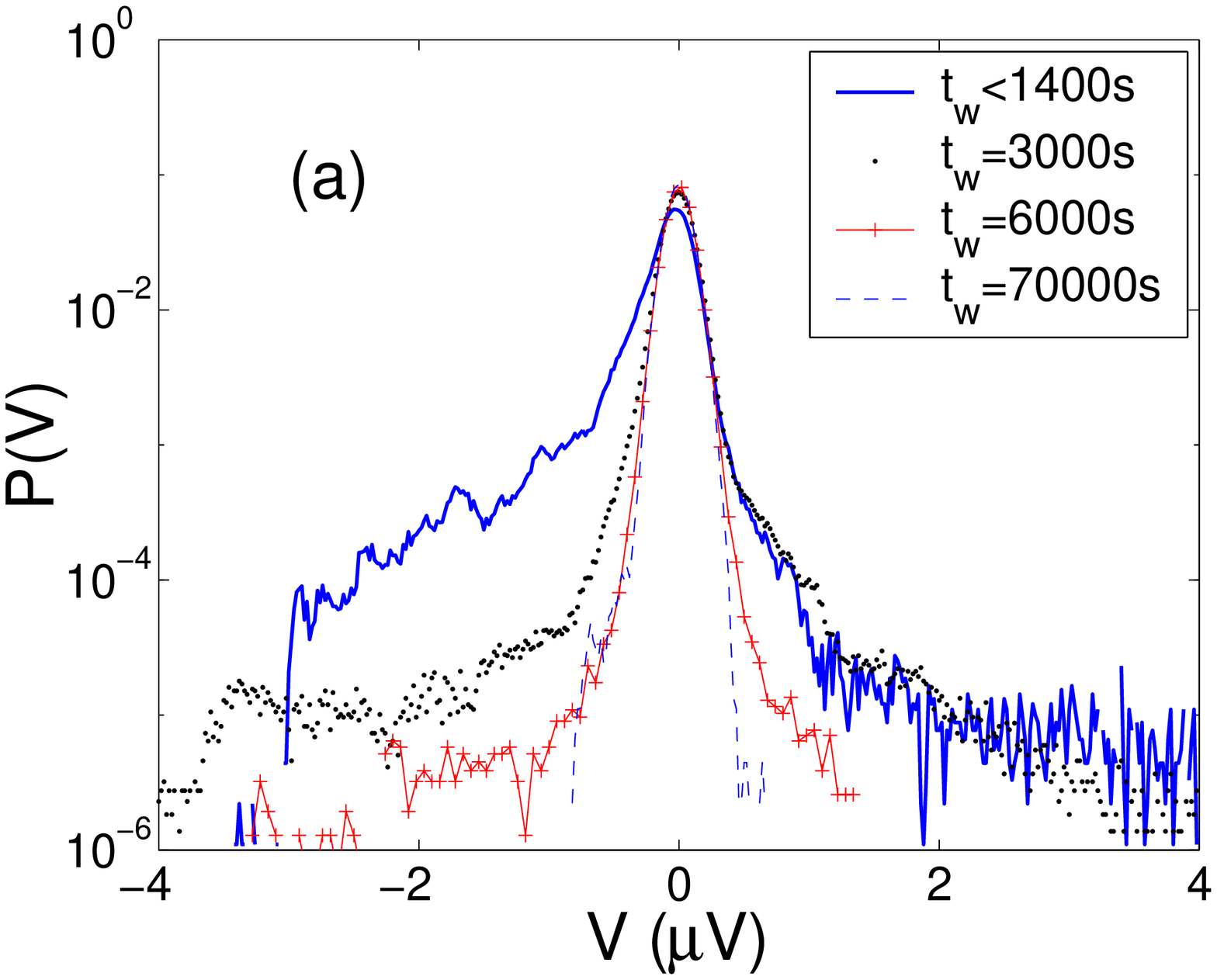}}
{\epsfxsize=0.5\linewidth
\epsffile{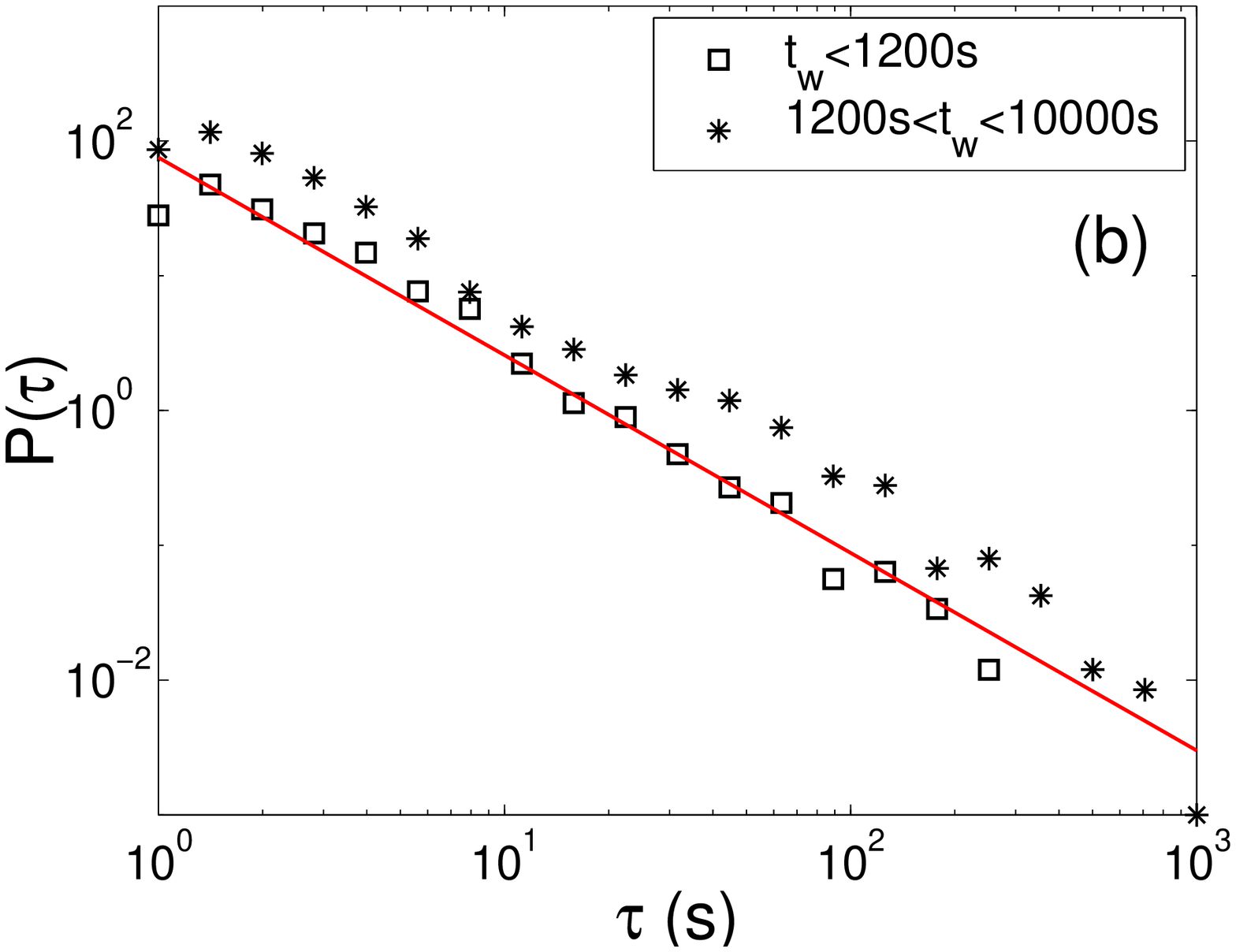}}} \caption{ {\bf PDF of
voltage noise in polycarbonate} Typical PDF of the noise signal of
polycarbonate measured at various  $t_w$  }
 \label{PDFpolyca}
\end{figure}

To further characterize the observed dynamics we have computed the
probability density function (PDF) of the signals, which  is
plotted in fig.\ref{PDFpolyca}(a) for different $t_w$. We clearly
see that the PDF, measured at small $t_w$, has very high tails
which become smaller and smaller at large $t_w$. Finally the
Gaussian profiles is recovered after $24h$. In a more quantitative
way we find that the Kurtosis $Ku$ (which is 0 for a Gaussian) of
the distribution  is a decreasing function of time  which can be
roughly approximated by:
    $  Ku= (6\pm 1) \ (t_w/1000)^{-0.55 \pm 0.05)}$.
 The time interval $\tau$
between two successive pulses is  power law distributed.  In order
to study the distribution $P(\tau,t_w)$ of $\tau$ we have first
selected the signal fluctuations with amplitude larger than a
fixed threshold, which has been chosen between 3 and 4 standard
deviations of the equilibrium noise,i.e. the noise predicted by
FDT. We have then measured the time intervals $\tau$ between two
successive large fluctuations. The PDF $P(\tau,t_w)$ computed for
$t_w<20min$ and for $20min<t_w<3h$ is plotted in
fig.\ref{PDFpolyca}b). We clearly see that $P(\tau,t_w)$ is a
power law, specifically $P(\tau)\propto {1 \over \tau^{1+\mu}}$
with
 $\mu \simeq 0.4\pm 0.1$.
This result agrees with one of the predictions  of the trap
model,\cite{trap}, which presents non trivial violation of FDT
associated to an intermittent dynamics. In the trap model $\tau$
is  a power law distributed quantity with an exponent $1+\mu$
that, in the glass phase, is smaller than 2.  However there are
important differences between the dynamics of our system and that
of the trap model. Indeed in this model one finds short and large
$\tau$ for any $t_w$ which is in contrast with our system because
the probability of finding short $\tau$ seems to decrease as a
function of $t_w$. But this effect could be a consequence of the
imposed threshold. It seems that there is no correlation between
the $\tau$ and the amplitude of the associated bursts. Finally the
maximum distance $\tau_{max}$ between two successive pulses grows
as function $t_w$ logarithmically, that is $\tau_{max}=[10+ 152\
\log(t_w/300)] s$ for $t_w>300s$. This slow relaxation of the
number of events per unit time  shows that the intermittency is
related to aging.

In conclusion, we have observed a large violation of FDT in the
dielectric thermal noise of an aging polymer glass. Such a large
violation is produced by rare events of high amplitude. The
important question is now to understand the physical origin of
these big events in the electrical thermal noise of the sample. A
possibility is the triboelectricity produced by the aging induced
stress relaxation in the sample. Mechanical and acoustical
measurements performed in parallel with dielectric measurements
could clarify this problem. Work is in progress.

We acknowledge useful discussion with and L. Bellon, L.
Cugliandolo and J.P. Bouchaud. We thank P. Metz and F. Vittoz for
technical support. This work has been partially supported by the
R\'egion Rh\^one-Alpes contract ``Programme Th\'ematique :
Vieillissement des mat\'eriaux amorphes''.



\end{document}